\documentclass[%
 aip,
 amsmath,amssymb,
 reprint,%
]{revtex4-1}

\usepackage{graphicx}
\usepackage{dcolumn}
\usepackage{bm}

\usepackage[utf8]{inputenc}
\usepackage[T1]{fontenc}
\usepackage{mathptmx}

\begin{document}

\preprint{AIP/123-QED}

\title[Continuous scanning of a dissipative Kerr-microresonator soliton comb for broadband, high resolution spectroscopy]{Continuous scanning of a dissipative Kerr-microresonator soliton comb for broadband, high resolution spectroscopy}

\author{N. Kuse}
\altaffiliation[Currently at ]{Institute of Post-LED Photonics, Tokushima University.}
\email{kuse.naoya@tokushima-u.ac.jp}
\affiliation{ 
IMRA America Inc., Boulder Research Labs, 1551 South Sunset St, Suite C, Longmont, CO, 80501, U.S.A
}
\author{T. Tetsumoto}
\affiliation{ 
IMRA America Inc., Boulder Research Labs, 1551 South Sunset St, Suite C, Longmont, CO, 80501, U.S.A
}
\author{G. Navickaite}
\affiliation{ 
LIGENTEC SA, EPFL Innovation Park L, Chemin de la Dent-d'Oche 1B, Switzerland  CH-1024 Ecublens, Switzerland
}
\author{M. Geiselmann}
\affiliation{ 
LIGENTEC SA, EPFL Innovation Park L, Chemin de la Dent-d'Oche 1B, Switzerland  CH-1024 Ecublens, Switzerland
}
\author{M. E. Fermann}
\affiliation{ 
IMRA America Inc., 1044 Woodridge Ave, Ann Arbor, MI, 48105, U.S.A
}

\date{\today}

\begin{abstract}
Dissipative Kerr-microresonator soliton combs (hereafter called soliton combs) are promising to realize chip scale integration of full soliton comb systems providing high precision, broad spectral coverage and a coherent link to the micro/mm/THz domain with diverse applications coming on line all the time. However, the large soliton comb spacing hampers some applications. For example, for spectroscopic applications, there are simply not enough comb lines available to sufficiently cover almost any relevant absorption features. Here, we overcome this limitation by scanning the comb mode spacing by employing PDH locking and a microheater on the microresonator, showing continuous scanning of the soliton comb modes across nearly the full FSR of the microresonator without losing soliton operation, while spectral features with a bandwidth of as small of 5 MHz are resolved. Thus, comb mode scanning allows to cover the whole comb mode spectrum of tens of THz bandwidth with only one chip-scale comb.
\end{abstract}

\maketitle

%

\section{Introduction}
Optical frequency combs generated from high-Q microresonators with a pump cw laser have been attracting significant interest for potential chip scale integration in the past decade especially after the experimental discovery of dissipative Kerr microresonator soliton combs \cite{Kippenberg_review18, gaeta2019photonic}. Soliton combs are in a mode-locked state, producing highly coherent, ultrashort pulses, with low noise comb spectra \cite{Herr_soliton}. Soliton combs have been applied to coherent optical communication \cite{Kippenberg_communication17}, low noise RF generation \cite{OEwaves_RF15}, ultra-fast ranging \cite{Kippenberg_distance18, Vahala_distance18}, optical frequency synthesizers \cite{Papp_synthesizer18}, to name a few. 

Another significant advantage of soliton combs in addition to chip scale is the large comb mode spacing, enabling easy access to each comb mode. Accessing each comb mode is very difficult for conventional small comb spacing combs (100 MHz – 1 GHz), e.g. based on mode-locked fiber laser combs. Conventionally, one indirectly accesses each comb mode by using the multi-heterodyne technique, i.e. dual-comb technique, in which two mutually phase locked fiber frequency combs with slightly different comb mode spacings are interfered \cite{coddington2008coherent, coddington2016dual}. Spectroscopy based on the dual-comb technique (called dual-comb spectroscopy) has been considered as one of the most powerful broadband, high speed spectroscopic tools. In addition, in conjunction with (small) optical comb mode shifting, dual-comb spectroscopy can provide high resolution \cite{hsieh2014spectrally, okubo2015near}. However, dual-comb systems are very complicated and expensive. By utilizing soliton combs, we can replace dual-comb systems with a single soliton comb, providing a chip-scale and low-cost spectroscopic tool with simultaneous broadband and high speed capability. In a comb-resolved system enabled by a soliton comb, soliton comb modes can be easily spatially-resolved by using a spectral disperser with moderate spectral resolution and converted to the electrical domain by photo detectors as conceptually shown in Fig. 1. In Fig. 1, an arrayed waveguide grating is used as a spectral disperser, followed by photo detectors and analog –digital converters, which are integrated in a Si platform (Note that the arrayed waveguide grating can be also integrated in the same as the microresonator). The comb-resolved system is useful not only for spectroscopy, but also for 3D imaging \cite{hase2018scan}, bio \cite{michaud2012whispering} and fiber optic sensors \cite{kuse2013static}, where each comb mode is used as spatially-separated, multiple cw lasers (coherence between the comb modes is also advantageous for some applications). However, soliton combs generally provide only very sparse spectral information, which is typically not enough for most spectroscopy applications. The sparsity is determined by the comb mode spacing of the soliton comb, which is typically 100 GHz – 1 THz for microresonators based on CMOS compatible platforms such as Si$_3$N$_4$ \cite{Kippenberg_SiN16, Gaeta_heater, Papp_OL_SSB18}.

In principle, the sparsity can be reduced via frequency shifting of the soliton comb modes, as typically implemented in high resolution dual-comb spectroscopy. However, for microcombs scanning of the comb modes is not as straight-forward as with conventional combs. A first challenge is that detuning between the cw pump laser and one of the resonances of the microresonator has a very limited allowable range in which soliton combs are maintained. Thus, both the cw pump laser and the resonance have to be scanned simultaneously by almost the same amount, which prohibits large and complex scanning of the soliton comb modes. A second challenge is that soliton combs require a much larger scan range than conventional combs to cover any optical frequency in-between the soliton comb modes. For example, soliton combs around 1550 nm (roughly 200 THz) with 100 GHz comb spacing require 0.05 \verb 

In this letter, we demonstrate large range, continuous scanning of a soliton comb by overcoming the first challenge, combined with the use of a microheater for the second challenge. A feedback loop based on Pound-Drever-Hall (PDH) locking is employed to fix the detuning between the cw pump laser and a resonance. The detuning is fixed even when the soliton comb modes are scanned, i.e. when the microresonator cavity resonance is scanned, the cw pump laser automatically follows the resonance frequency change. When PDH locking is engaged, a microheater on the microresonator enables continuous scanning with a scan range > 190 GHz without losing soliton operation. To demonstrate the feasibility of high precision spectroscopy via comb mode scanning, the system is applied to the characterization of H$^{13}$C$^{14}$N (hereinafter HCN) in the gas phase as well as resolving the modes of a fiber cavity. Comb mode scanning improves the frequency resolution by > 50000 times, compared with the case when no comb mode scanning is involved (280 GHz). Spectral features of as small as 5 MHz, only limited by the absorption linewidth of a test sample, is obtained.
\begin{figure}[!ht]
\centering
\fbox{\includegraphics[width=\linewidth]{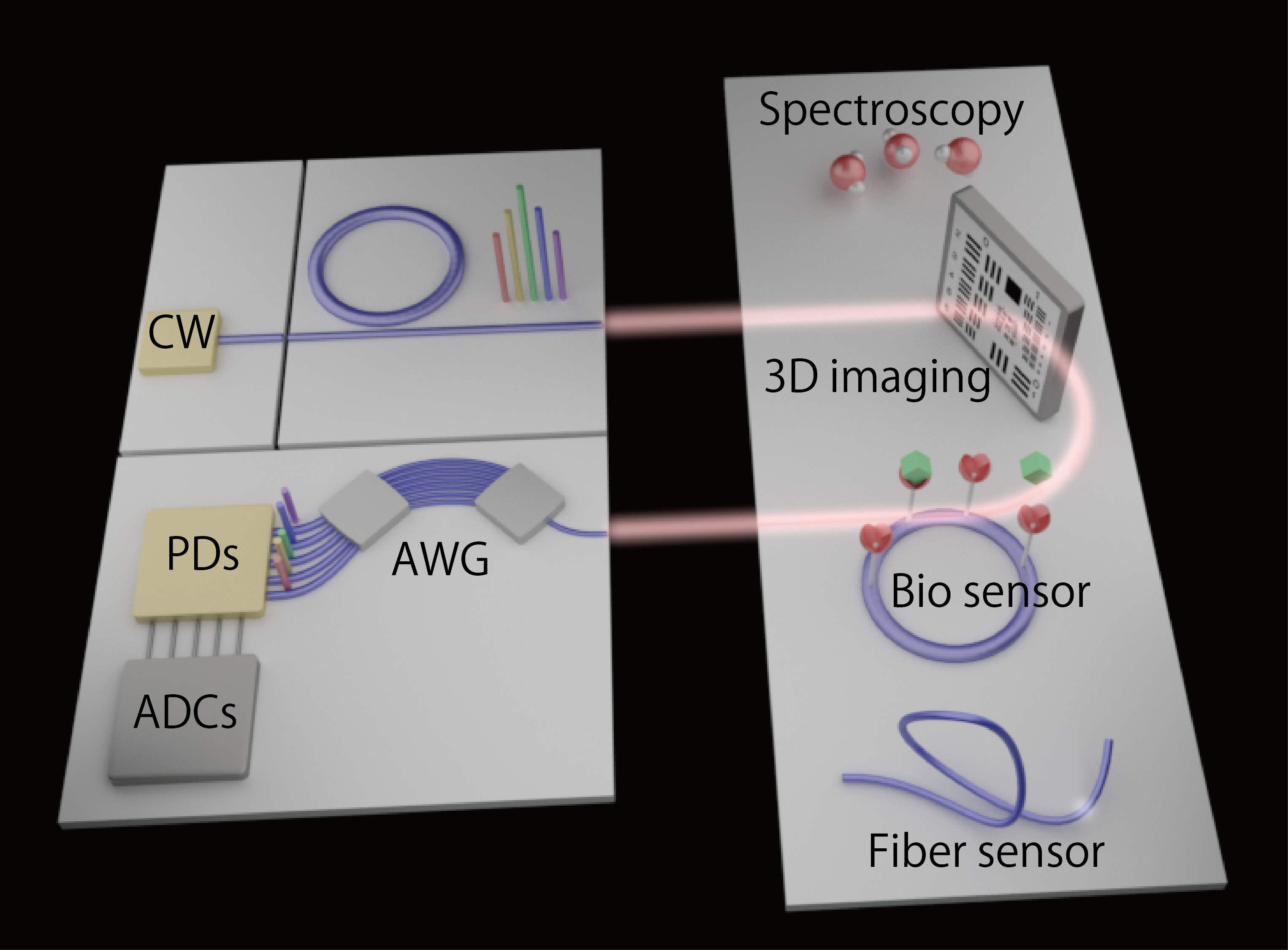}}
\caption{ Conceptual setup for a chip-scale spectroscopic system. Because chips would be based on different platforms, butt-coupling, photonic wire bonding, or other technologies would be required to connect the chips. Depending on applications, the soliton comb does not need to propagate in free space. PDs, photo detectors; AWGs, arrayed waveguide gratings; ADCs, analog-digital converters.}
\end{figure}  

\section{Result}

\subsection{Experimental schematic}
Conventionally, PDH locking is used to lock a cw laser to a cavity, where the optical frequency of the cw laser is locked to one of the cavity resonances \cite{black2001introduction}. Very recently, PDH locking has been utilized for preserving soliton comb operation for a long time, where detuning between the cw pump laser and one of the resonances of the microresonator has been fixed \cite{Papp_PRL18}. The detuning is determined by the RF frequency applied to a phase modulator, i.e. a blue-side sideband from the cw laser is locked to a resonance frequency of the microresonator. In this letter, PDH locking is used for soliton comb mode scanning for the first time. When the relevant sideband is locked to the resonance, soliton comb modes can be scanned without losing soliton comb operation because of the fixed detuning as shown in Fig. 2(a). An electric current is applied to a microheater on a microresonator to heat the microresonator, changing the resonance frequency of the microresonator. Because of PDH locking, the optical frequency of the cw laser is autonomously changed during scanning of the resonance frequency of the microresonator, keeping the detuning fixed.

An experimental setup is shown in Fig. 2(b).  A cw laser around 1550 nm is directed through a phase modulator and dual-parallel Mach-Zehnder modulator (DP-MZM). The phase modulator is driven by an RF oscillator with modulation frequency and power of 1.3 GHz and -20 dBm, respectively. The modulation frequency determines the detuning. The modulation power is weaker than typically implemented for PDH locking to ensure the sideband for PDH locking does not affect the cavity dynamics in the microresonator. The DP-MZM operated in a carrier-suppressed single sideband mode is used to initiate a soliton comb \cite{Papp_OL_SSB18, kuse2019control}. By quickly applying a DC voltage to a VCO, the pump frequency can be quickly shifted by a few GHz / 20 ns from blue to red wavelengths. The fast frequency shifting beyond the thermo-optic response in the Si$_3$N$_4$ (SiN) microresonator enables the generation of a single soliton comb with high fidelity. An output from the DP-MZM is amplified by an erbium-doped fiber amplifier (EDFA) up to a power of 500 mW, and coupled into a microresonator. The FSR and loaded Q of the SiN microresonator are about 280 GHz and 10$^6$, respectively. An aluminum microheater is installed on the SiN microresonator separated by a SiO$_2$ cladding layer. The transmission of the cw laser is photo-detected and mixed with the same RF oscillator used for the phase modulator, generating an error signal via demodulation. The error signal is signal-processed by PI filters and fed back to both a VCO for the DP-MZM and the cw laser. Although, in principle, feeding back only to the cw laser works, we have to modulate both the VCO and the cw laser, because the cw laser has a minimum discrete frequency step, while the VCO can be tuned continuously. Alternatively, feeding back to the microheater can also work while changing the frequency of the cw laser, but in this case the feedback loop is nonlinear since the frequency response of the microheater is proportional to the square of the applied voltage. The generated soliton comb is used for a spectroscopy experiment and the optical spectrum of the soliton comb is monitored by an optical spectrum analyzer when the soliton comb modes are scanned. 

\begin{figure}[!ht]
\centering
\fbox{\includegraphics[width=\linewidth]{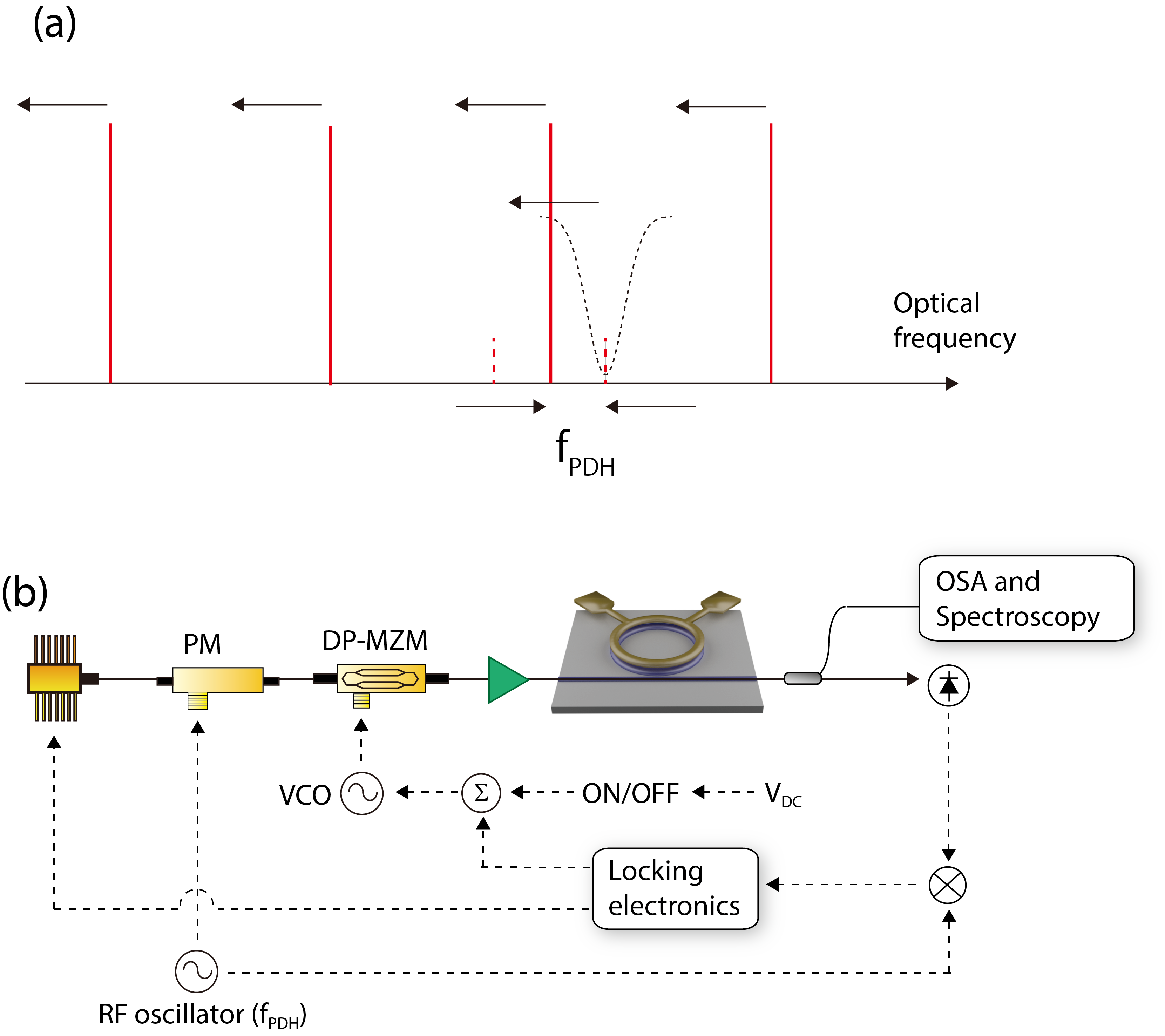}}
\caption{(a) Schematic of working principle. One of the sidebands generated from a phase modulator (blue side) is locked to a resonance of the microrensonator. The detuning is determined by the RF frequency for PDH locking. (b) Schematic of experimental setup. A microheater is deposited on a SiN microresonator. PM, phase modulator; VCO, voltage controlled oscillator; OSA, optical spectrum analyzer.}
\end{figure}

\subsection{Single soliton comb and PDH locking}
Figure 3(a) shows the optical spectrum of a single soliton comb. The comb modes are clearly resolved, showing about 280 GHz comb mode spacing and a smooth envelope. After generating the soliton comb, the pump frequency, which is located on the red-side of a resonance shifted by about 1.3 GHz, is modulated to check an error signal against detuning as shown in Fig. 3(b). The error signal has a zero crossing when the sideband generated from the phase modulator becomes equal to the resonance frequency. Figure 3(c) shows an error signal when the feedback loop is off and on. When the feedback loop is off, the error signal is fluctuating, originating from the fluctuations of the pump frequency and the resonance frequency. Once the feedback loop turns on, the detuning between pump and resonance frequency is fixed at the RF frequency applied for the phase modulator ($\sim$ 1.3 GHz), showing no fluctuations of detuning. Because of fixed detuning, the soliton comb can be maintained for more than one day as shown in Fig. 3(d). We could operate the soliton comb indefinitely, if coupling to the microresonator would have sufficient long-term stability.

\begin{figure}[!ht]
\centering
\fbox{\includegraphics[width=\linewidth]{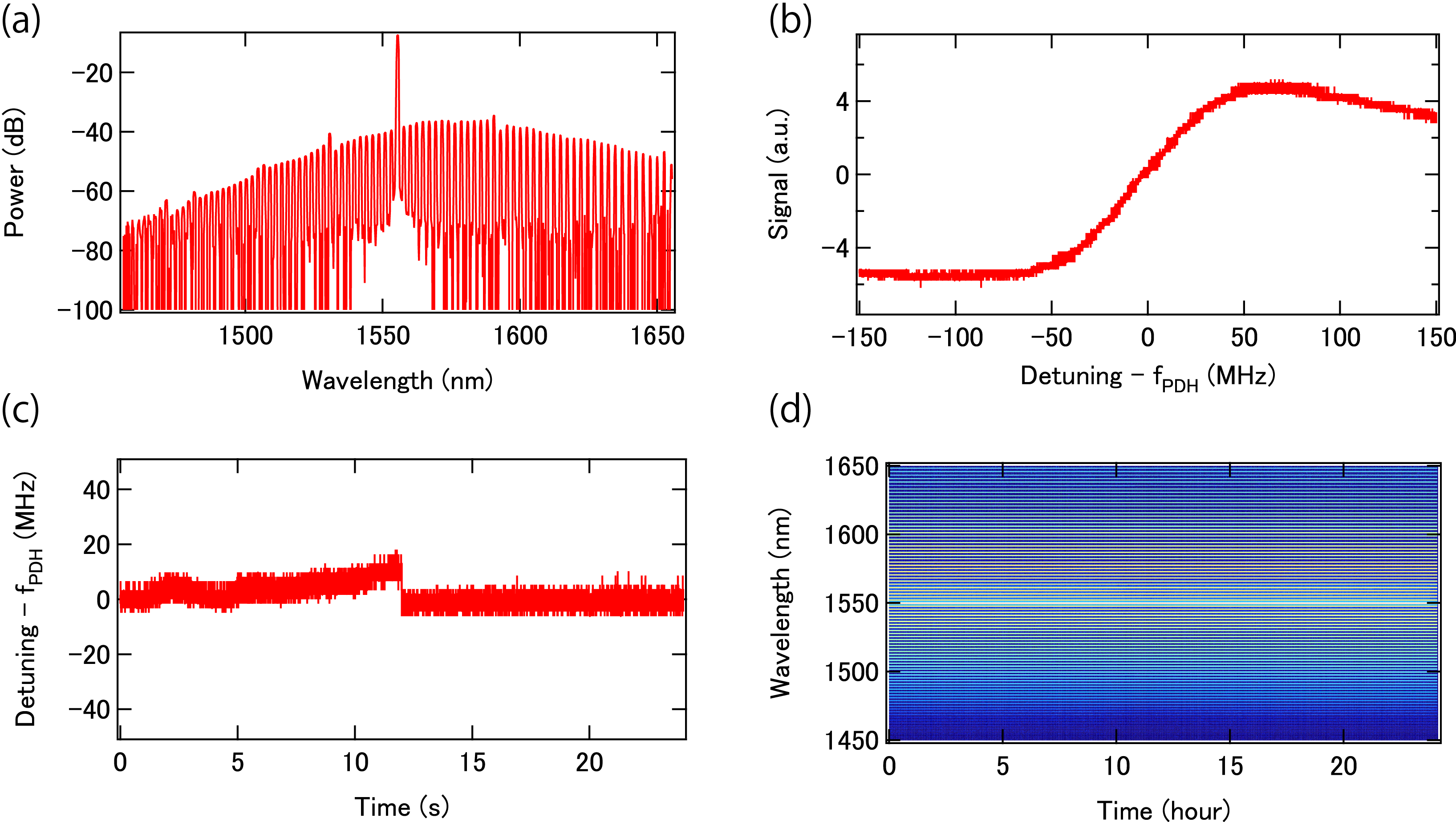}}
\caption{(a) Optical spectrum of the soliton comb. (b) Error signal of PDH locking. (c) Detuning with (> 11 s) and without (< 11 s) turning on the feedback loop via PDH locking. (d) Optical spectrum of the soliton comb when the feedback loop is closed.}
\end{figure}

\subsection{Soliton comb mode scan}
When the soliton comb is scanned, an electric current is applied to the microheater. The temperature of the microheater is increased by Joule heat generated from applied electric power ($I^2R$, where $I$ and $R$ are electric current and resistor, respectively). Figure 4(a) and (b) show an optical spectrum of the soliton comb at different spectral spans. Because of PDH locking, the soliton comb is not lost even during scanning. As emphasized in Fig. 4(b), the soliton comb modes are scanned by about 1.5 nm (i.e. 190 GHz) with about 200 mW electric power, which corresponds to 69 \verb 

\begin{figure}[!ht]
\centering
\fbox{\includegraphics[width=\linewidth]{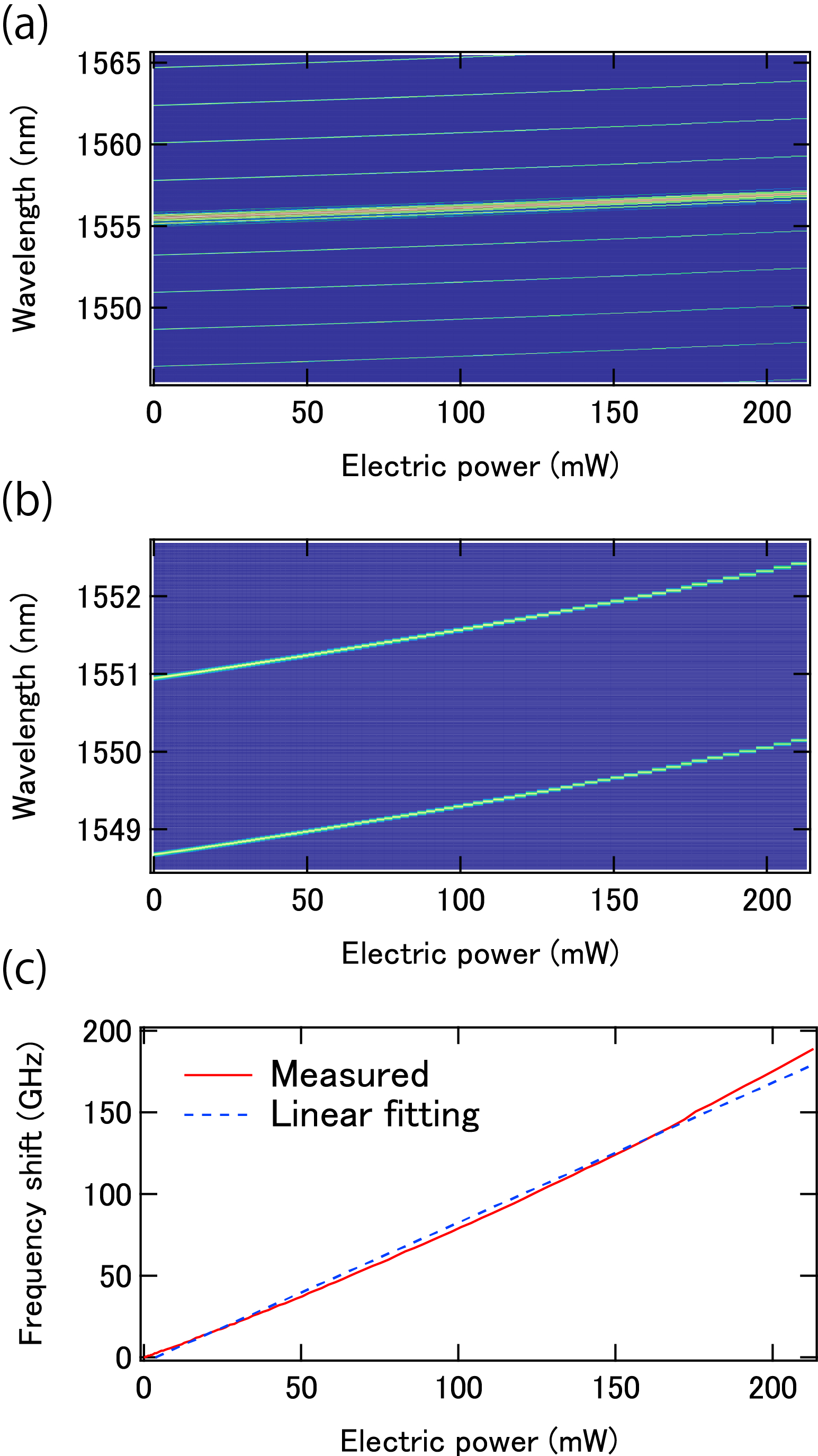}}
\caption{(a) Optical spectrum of the soliton comb when Joule heat is generated from applied electric power. (b) Magnified optical spectrum of (a). (c) Frequency shift of one of the soliton comb modes.}
\end{figure}

\subsection{Spectroscopy}
As a proof-of-concept experiment for broadband spectroscopy, gas-phase H$^{13}$C$^{14}$N (hereafter HCN) is used as a sample. HCN has many absorption lines in the C-band \cite{swann2005line}. Instead of measuring all comb modes at one time by using a WDM as shown in Fig. 1, only one comb mode is measured at one time because of the limited equipment in our lab. The separated single comb mode, which is continuously frequency-scanned, is photo-detected and digitized by an oscilloscope (RIGOL, DS1054Z). We measured three absorption lines 1532, 1541, and 1555 nm, which are allocated as R15, R0, and P17, respectively. Since scanning is nonlinear against the applied electric current/voltage, which means optical frequency of the soliton comb modes is scanned nonlinearly, the nonlinearity is compensated by monitoring optical frequency variations via an imbalanced fiber Mach-Zehnder modulator. The transmissions of the absorption lines are shown in Fig. 5. The transmissions show clearly resolved Voigt profiles (blue dotted curves in Fig. 5) with spectral FWHM width of about 1.5 GHz, corresponding to the expected theoretical curve from environmental temperature and pressure. 

\begin{figure}[!ht]
\centering
\includegraphics[width=\linewidth]{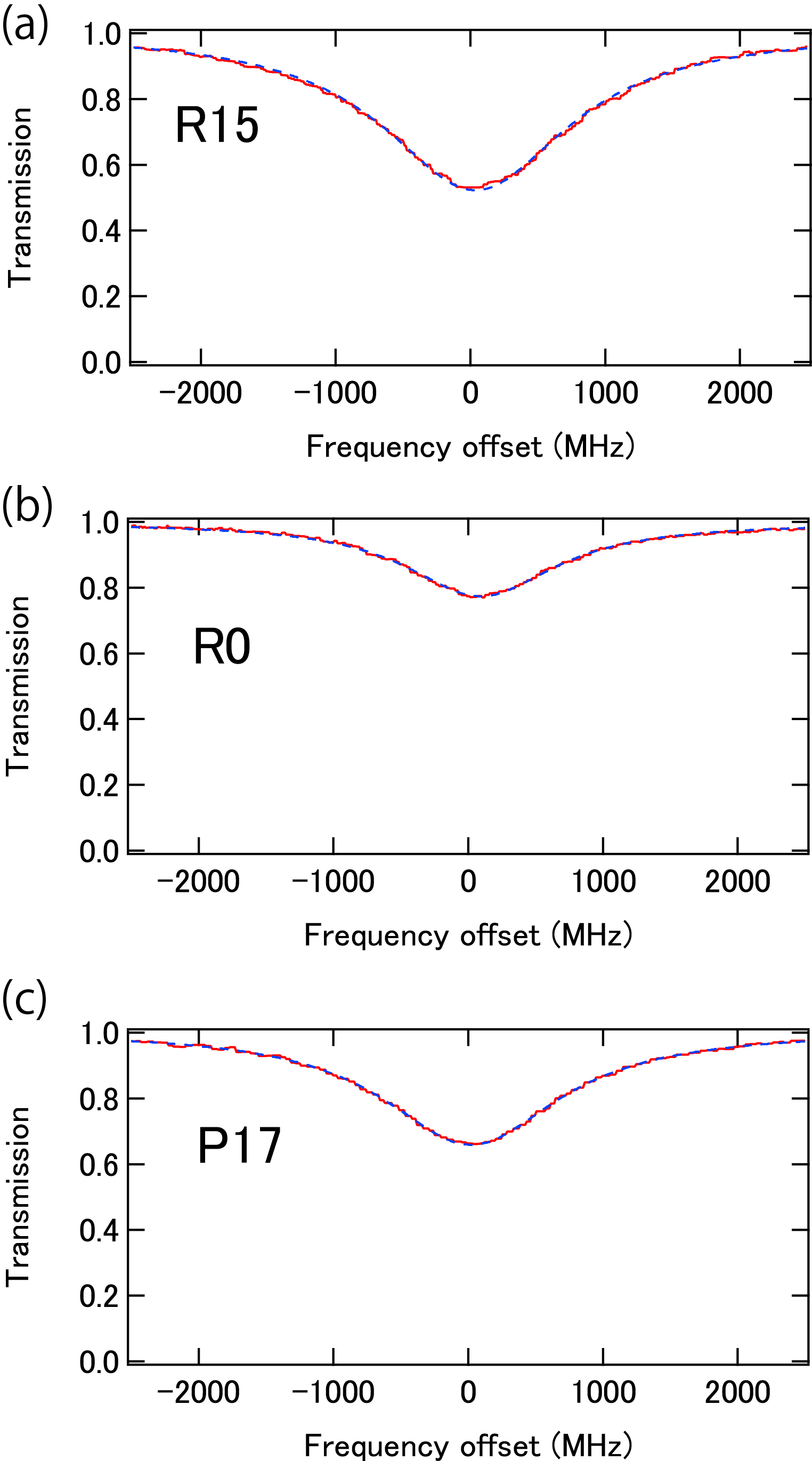}
\caption{Transmission optical spectra from a gas-phase HCN at R15(a), R0 (b), and P17 (c). The red curves show experimental results and the dotted blue curve show expected theoretical curves calculated from environmental temperature and pressure.}
\end{figure}

To demonstrate higher resolution spectroscopy, we also use a fiber resonator as a sample. The fiber resonator consists of a 1.5 m fiber coupler with 90: 10 splitting ratio. As shown in Fig. 6(a), narrow absorption lines are measured every $\sim$ 130 MHz. Figure 6(b) shows an expanded absorption line. The absorption line has a FWHM of as small as 5 MHz, which is the narrowest absorption line measured by soliton combs. The ultimate resolution is determined by the instantaneous linewidth of the comb modes, which is on the order of kHz for a compact external cavity diode laser.

Although the measurements presented here are based on intensity sensitive absorption spectroscopy, phase sensitive absorption spectroscopy for high precision and sensitivity can be realized by wavelength/frequency modulation spectroscopy \cite{bomse1992frequency}, which has been established in spectroscopic systems based on a tunable single longitudinal mode cw laser. The frequency accuracy can be also very high, combined with an external reference or detection of the carrier envelope offset frequency while counting the frequency of the comb mode spacing. When direct detection of the comb mode spacing is difficult, frequency down conversion through sideband generation via electro-optic modulation can be implemented \cite{Kippenberg_SiN16, xue2016thermal}. However, quantitative analysis and demonstration are beyond the scope of this report.

\begin{figure}[!ht]
\centering
\fbox{\includegraphics[width=\linewidth]{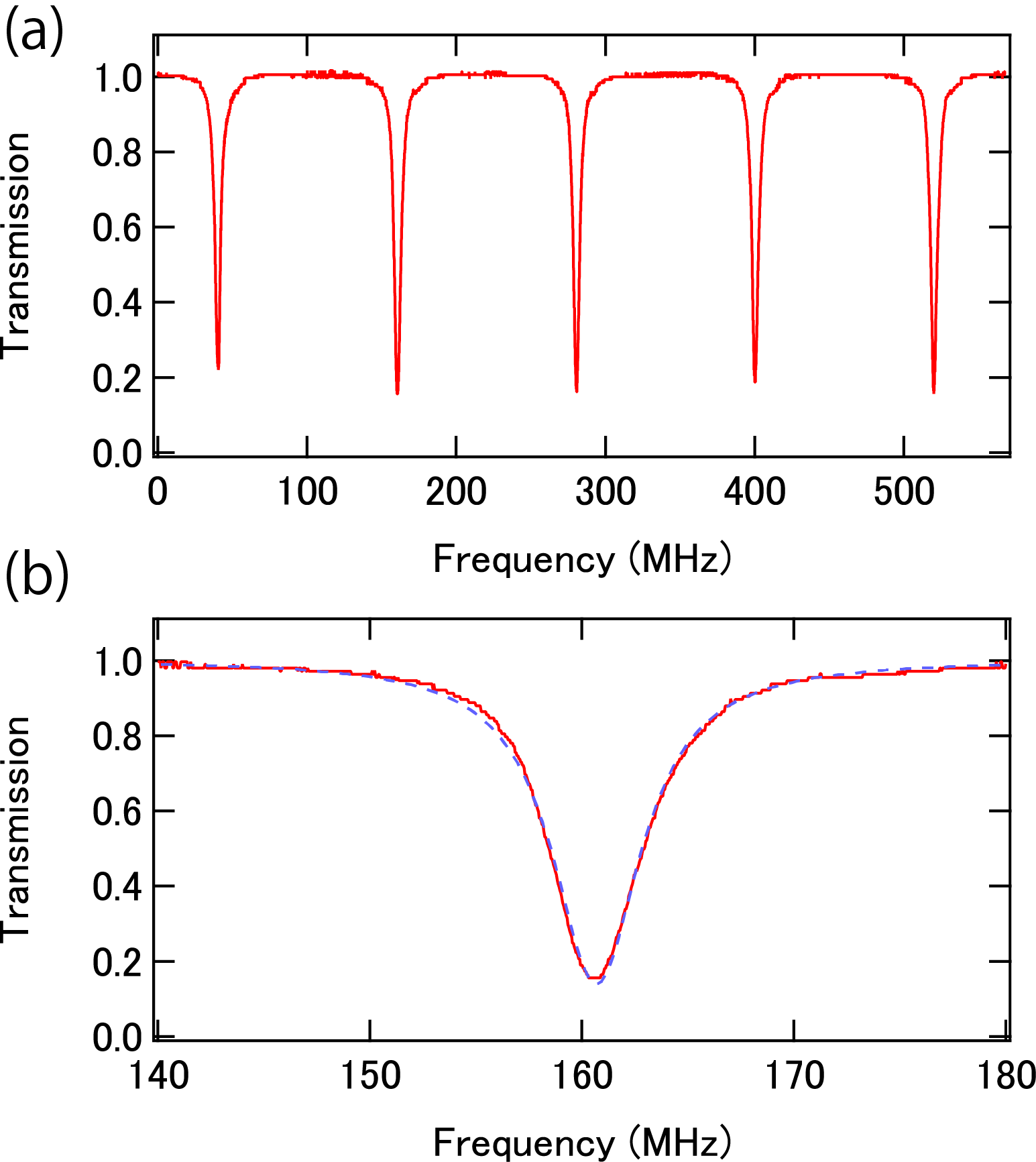}}
\caption{(a) Transmission optical spectra from the fiber resonator. (b) Magnified spectrum of (a). Dotted blue curve is Lorentz curve.}
\end{figure}

\section{Discussion and conclusion}
The achieved scan range here did not reach one FSR, as would be desired for an actual spectroscopic instrument due to the limited current that could be applied to the heaters. By using microcombs with a smaller comb spacing such as 100 GHz, comb mode scanning of more than one FSR can be realized \cite{Kippenberg_chip18}. Furthermore, although the width of the microheater is designed to be the same as the microresonator, a larger microheater width would increase the scan range by reducing the probability of electromigration, which is proportional to the inverse of the cross section. This however goes with a drop in resistance and increase in electrical power. Alternatively, microheaters made by a different material such as Pt would also allow a larger scan range by reducing the effect of electromigration \cite{Gaeta_heater, yu2018gas}. With these small modifications, comb mode scanning of more than 200 GHz, which is larger than the comb mode spacing of most microcombs, can be envisioned, providing for full spectral coverage of the comb spectrum. 

At present, the scan speed is slow (> 1 s for Fig. 4) and limited by the scan speed of our bulky cw pump laser. However, much higher scan speeds can in principle be demonstrated. Ultimately, the scan speed is limited by the modulation speed of the microheaters or the cw pump laser, which can be < 1 ms. To reach a scan speed limited by microheaters, distributed feedback lasers (DFB lasers) can be used at the cost of the broader linewidth of DFB lasers ($\sim$  1 MHz). Alternatively, self-injection locked semiconductor lasers could be also used for scanning applications \cite{Lipson_chip18, pavlov2018narrow}. 

The method can be applied for any wavelength in principle. Phase modulators for PDH locking from the visible to 2 $\mu$m are commercially available. For longer wavelength, where phase modulators are not currently available, direct modulation of cw pump lasers can be used to add phase modulation.

In conclusion, we demonstrated continuous scanning of a dissipative Kerr soliton comb from a microresonator, in which PDH locking was used to fix the detuning between a cw pump laser and a resonance frequency of the microresonator, allowing for a scan range as large as 190 GHz (and potentially by more than the FSR for most soliton combs). In addition, proof-of-concept experiments of broadband, high resolution spectroscopy were demonstrated, showing a resolution of as high as 5 MHz. The spectroscopic system can be fully integrated in silicon photonics, in conjunction with soliton combs from high Q microresonators, integrated cw lasers, electro-optic modulators, AWGs, and PDs. Scanning soliton microcombs thus afford a unique pathway to the realization of practical, mass producible, spectroscopic tools with superior sensitivity, potentially fast measurement, and broader spectral coverage than currently available. Moreover, since comb mode scanning as presented here is mode-hop free, the system is essentially equivalent to providing multiple coherently-frequency-modulated wavelength channels, with potentially many other applications, for example related to laser ranging, radar, 3D imaging and general signal processing.

\section*{Supplementary material}
See supplementary material for a mathematical expression for more details about experimental setups.

\begin{acknowledgments}
We acknowledge Yi Xuan and Cong Wang in Purdue University at the beginning of the experiment.
\end{acknowledgments}

\appendix

\nocite{*}
\bibliography{aipsamp}

\end{document}